\documentclass[sigconf]{acmart}
\AtBeginDocument{%
  }

\setcopyright{acmlicensed}
\copyrightyear{2025}
\acmYear{2025}
\acmDOI{XXXXXXX.XXXXXXX}
\acmConference[2025 ACM SIGMOD/PODS]{ACM SIGMOD/PODS International Conference on Management of Data}{June 22-27, 2025}{Berlin, GER}
\acmISBN{XXX-X-XXXX-XXXX-X/XXXX/XX}

\begin{document}

\title[Trustworthy LLM-Driven Process Modeling, Prediction and Automation]{From Theory to Practice: Real-World Use Cases on Trustworthy LLM-Driven Process Modeling, Prediction and Automation}

\author{Peter Pfeiffer}
\email{peter.pfeiffer@dfki.de}
\affiliation{\institution{German Research Center for Artificial Intelligence (DFKI) and Saarland University}
  \city{66123 Saarbrücken}
  \state{Saarland}
  \country{Germany}
}

\author{Alexander Rombach}
\email{alexander\_michael.rombach@dfki.de}
\affiliation{%
  \institution{German Research Center for Artificial Intelligence (DFKI) and Saarland University}
  \city{66123 Saarbrücken}
  \state{Saarland}
  \country{Germany}
}

\author{Maxim Majlatow}
\email{maxim.majlatow@dfki.de}
\affiliation{%
  \institution{German Research Center for Artificial Intelligence (DFKI) and Saarland University}
  \city{66123 Saarbrücken}
  \state{Saarland}
  \country{Germany}
}

\author{Nijat Mehdiyev}
\email{nijat.mehdiyev@dfki.de}
\affiliation{%
  \institution{German Research Center for Artificial Intelligence (DFKI) and Saarland University}
  \city{66123 Saarbrücken}
  \state{Saarland}
  \country{Germany}
}

\renewcommand{\shortauthors}{Pfeiffer, Rombach, Majlatow and Mehdiyev}

\begin{abstract}
Traditional Business Process Management struggles with rigidity, opacity, and scalability in dynamic environments, while emerging Large Language Models (LLMs) present transformative opportunities alongside risks. This paper explores four real-world use cases that demonstrate how LLMs, augmented with trustworthy process intelligence, redefine process modeling, prediction, and automation. Grounded in early-stage research projects with industrial partners, the work spans manufacturing, modeling, life-science, and design processes, addressing domain-specific challenges through human-AI collaboration. In manufacturing, an LLM-driven framework integrates uncertainty-aware explainable Machine Learning with interactive dialogues, transforming opaque predictions into auditable workflows. For process modeling, conversational interfaces democratize BPMN design. Pharmacovigilance agents automate drug safety monitoring via knowledge-graph-augmented LLMs. 
%For process modeling, conversational interfaces democratize BPMN design, while pharmacovigilance agents automate drug safety monitoring via knowledge-graph-augmented LLMs. 
Finally, sustainable textile design employs multi-agent systems to navigate regulatory and environmental trade-offs. We intend to examine tensions between transparency and efficiency, generalization and specialization, and human agency versus automation. By mapping these trade-offs, we advocate for context-sensitive integration — prioritizing domain needs, stakeholder values, and iterative human-in-the-loop workflows over universal solutions. This work provides actionable insights for researchers and practitioners aiming to operationalize LLMs in critical BPM environments.
\end{abstract}




\received{15.05.2025}
% \received[revised]{12 March 2025}
% \received[accepted]{5 June 2025}

\maketitle

% ... of the BPM lifecycle \cite{houy2010empirical} with phases \textit{strategy development} (1), \textit{definition and modeling} (2), \textit{implementation} (3), \textit{execution} (4), \textit{monitoring and controlling} (5), \textit{optimization and improvement} (6). 
\section{Introduction}
\label{sec:introduction}

The integration of Artificial Intelligence (AI) into Business Process Management (BPM) has long been recognized as a pathway to operational efficiency \cite{van2012process}. Nevertheless, conventional approaches — reliant on manual process modeling, rigid statistical methods, and non-transparent Predictive Process Monitoring (PPM) systems — struggle to address the dynamism and complexity of modern industrial workflows \cite{di2018predictive, maggi2014predictive}. While Process Mining (PM) has enabled organizations to visualize and analyze event logs, critical gaps persist: static models fail to adapt to real-time operational shifts, domain experts face technical barriers \cite{understanding, rizzi2020explainability}, and “black-box” AI systems weaken trust in high-stakes decision-making \cite{black_box, rudin2022interpretable,aliyeva2024uncertainty}. These limitations are particularly evident in industries such as manufacturing, healthcare, and sustainable design, where urgent customization, regulatory compliance, and multi-stakeholder collaboration demand flexible, transparent solutions \cite{mohseni2021multidisciplinary}.

Recent advances in Large Language Models (LLMs) and Explainable AI (XAI) signal a transformative shift, framing AI not just as an automation tool but as a collaborative partner in process management \cite{sebin2024exploring}. LLMs’ ability to interpret natural language, generate contextual insights, and interact with diverse data sources aligns with the evolving demands of BPM, where adaptability and human-AI collaboration are essential \cite{slack2023explaining}. However, realizing this potential requires addressing practical challenges: How can LLMs bridge the expertise gap between process modeling experts and domain specialists? Can AI-driven systems balance automation with accountability in regulated environments? What technical innovations are needed to scale these solutions across industries?

This paper addresses these questions through four real-world use cases, each stemming from recently funded research initiatives that explore distinct applications of LLMs in BPM along the BPM lifecycle \cite{Dumas2018_BPM}. Spanning manufacturing, cross-industry process modeling, drug safety monitoring, and sustainable textile design, these projects demonstrate how LLMs can support activities along the entire BPM lifecycle — from discovery and execution to monitoring and optimization — while tackling domain-specific challenges:
\begin{itemize}
    \item In \textbf{LLM-mediated interaction with trustworthy process predictions}, a framework integrating uncertainty-aware Machine Learning (ML), PM, and multi-agent LLMs transforms opaque predictions into auditable, interactive workflows. By grounding explanations in predictions from Manufacturing Execution System (MES) event logs and enabling natural language dialogues, the system empowers users to validate and adapt AI recommendations in real-time.
    \item For \textbf{conversational process modeling}, a chat-based interface allows users to model and refine Business Process Model and Notation (BPMN) models in an interactive manner, democratizing access to process design and optimization.
    %uses fine-tuned LLMs to translate natural language descriptions into executable BPMN models, democratizing access to process design.
    \item In \textbf{pharmacovigilance}, knowledge graph (KG)-augmented LLM agents support regulatory experts in drug safety monitoring by synthesizing unstructured reports and structured medical domain knowledge.
    %, streamlining medicines monitoring workflows to ensure safety.
    \item For \textbf{sustainable textile design}, a multi-agent LLM system navigates regulatory, environmental, and social trade-offs, empowering designers to align creative choices with circular economy principles.
\end{itemize}

Though distinct in scope, these cases share three transformative themes for BPM. First, they demonstrate \textit{human-centric AI}, where LLMs act as bridges between technical systems and non-expert users, whether through process visualizations or natural language queries. Second, they emphasize \textit{trustworthy automation} through XAI techniques — such as uncertainty quantification (UQ) and Retrieval-Augmented Generation (RAG) - that anchor LLM outputs in verifiable data sources. Third, they expose \textit{scalability and performance challenges} in integrating LLMs into legacy BPM infrastructures, from reconciling XML-based process models with generative AI to reducing errors in KG queries. By presenting these diverse perspectives, this paper contributes to next-generation process management in three ways. Practically, it shows how LLMs address industry pain points like manual process modeling bottlenecks. Methodologically, it highlights hybrid architectures that merge LLMs with traditional ML and participatory design. Theoretically, it challenges the view of LLMs as standalone solutions, advocating instead for their role in human-in-the-loop workflows.

The remainder of this paper is structured as follows: Sections~\ref{section_use_case_1}–\ref{section_use_case_4} detail the four use cases, emphasizing their technical architectures, LLM integration strategies, and (initial) empirical findings. Section~\ref{section_discussion} synthesizes challenges and opportunities, framing a research agenda for AI-driven BPM in the LLM era. Finally, \autoref{sec:conclusion} summarizes and concludes the paper.
\begin{figure*}[t]
  \centering
  \includegraphics[width=0.89\linewidth]{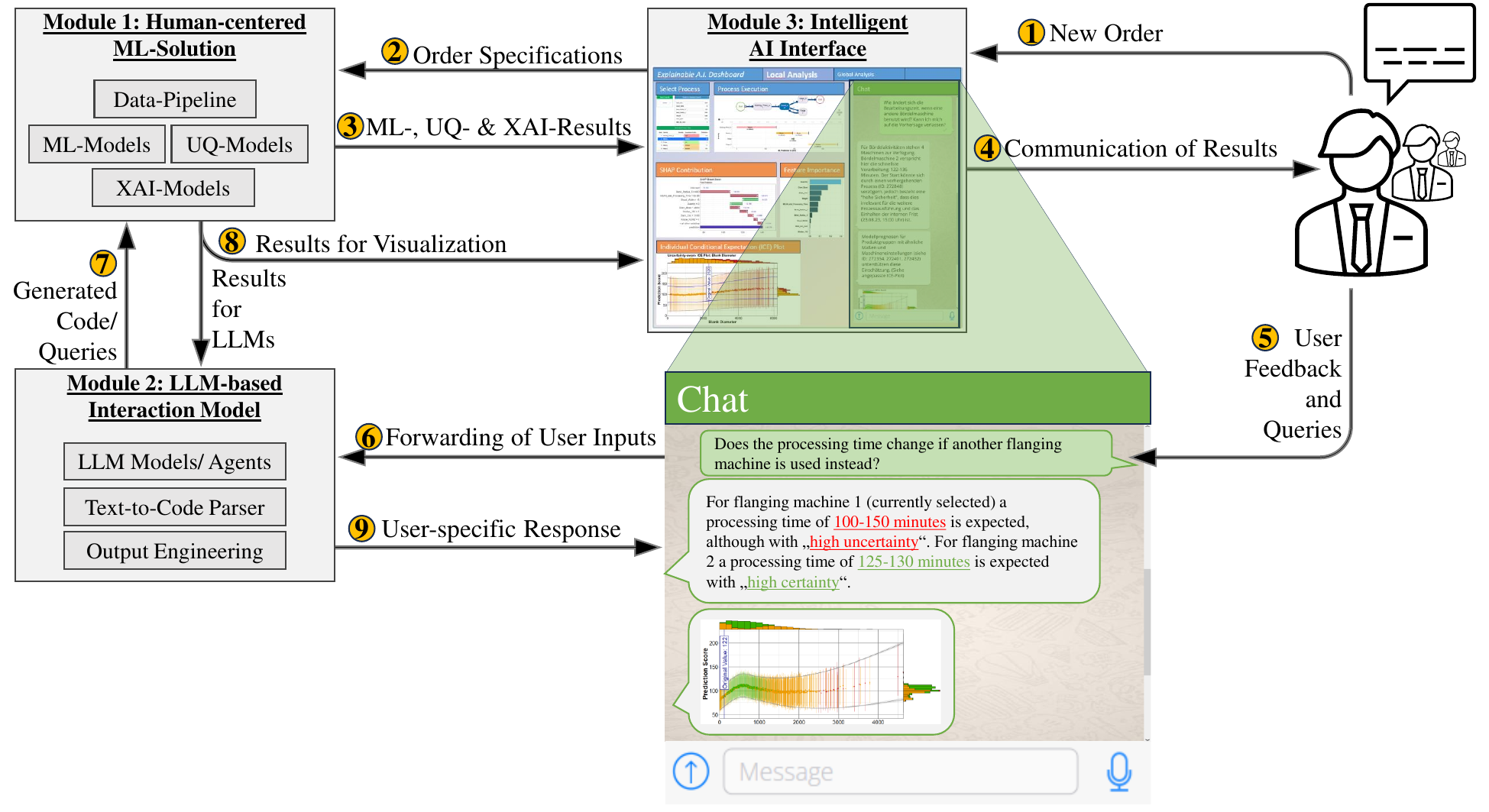}
  \Description{Conceptual Framework of Use Case 1: The diagram depicts a three‐module AI solution that integrates human‐centered ML, LLM–based interaction, and an intelligent AI interface. It orchestrates data processing, predictive analytics, UQ, XAI results, and user interaction. The goal is to provide a seamless workflow for production planning and execution, analyzing data and schedules with ML, UQ, and XAI techniques, and returning intelligible, user‐specific outcomes.}
  \caption{Three‐module AI solution that integrates human‐centered ML, LLM–based interaction, and an intelligent AI interface}
  \label{figure_use_case_1}
\end{figure*}

\section{Use Case 1 - LLM-Mediated Interaction with Trustworthy Process Predictions}
\label{section_use_case_1}
\subsection{Motivation}
Modern manufacturing systems generate vast volumes of event data through MES, capturing granular details of production machine states, operator actions, material flows, and quality checks \cite{mehdiyev2024counterfactual}. PM leverages these event logs to reconstruct and analyze production processes, offering insights into bottlenecks, inefficiencies, and deviations from planned schedules. While traditional PM excels at descriptive analytics, its capacity for predictive and prescriptive optimization remains limited, particularly in dynamic environments requiring real-time adaptation to urgent orders, material variability, or machine downtime \cite{teinemaa2019outcome}.

ML has emerged as a promising tool to augment PM in production scheduling and planning \cite{mehdiyev2024counterfactual}. PPM models, trained on historical event logs, can forecast production durations, optimize resource allocation, and simulate alternative workflows. However, their adoption introduces a critical challenge: the opacity of ML decision-making undermines trust in high-stakes scenarios. Production planners, tasked with validating schedules against contractual deadlines and quality standards, face a dilemma. They must reconcile ML’s predictive accuracy with the need for intuitive, auditable reasoning — a requirement unmet by "black-box" ML models that obscure the rationale behind their outputs.

Current XAI methodologies fail to resolve these barriers \cite{miller2023explainable}. Most approaches generate static explanations that do not align with the dynamic priorities of domain experts \cite{liao2020questioning}. This creates a critical bottleneck: machine-generated explanations lack the flexibility to accommodate human intuition, limiting users’ ability to interrogate, refine, or contextualize model outputs. Consequently, even accurate predictions risk rejection, perpetuating reliance on suboptimal traditional methods like expert estimations. The consequences of this gap are severe. Inaccurate or mistrusted predictions risk costly delays, contractual penalties, and resource underutilization. For example, overly conservative scheduling to hedge against opaque ML recommendations may idle machines or operators, while overly optimistic forecasts jeopardize on-time delivery.
\subsection{Approach}
% \subsubsection*{Detailed Description of Proposed Approach:}
The challenges outlined above, namely non-transparent ML-driven decision-making, misalignment between static explanations and human intuition, and the operational risks of mistrusted predictions, underscore a pressing need for frameworks that harmonize technical explainability with domain-specific interpretability. Traditional approaches treat explainability as a post-hoc justification layer, divorced from the dynamic, context-rich realities of production planning. To bridge this gap, we propose a process-driven human-AI collaboration framework that reimagines explainability as an iterative, process-aware dialogue. The conceptual framework, consisting of three modules, is shown in \autoref{figure_use_case_1}.

Central to this approach is the integration of PM, which grounds ML predictions in the temporal and causal dependencies captured by MES event logs. Module 1 collects and prepares event logs from various manufacturing steps, such as pressing, bending, or grinding. Supervised learning is applied to forecast processing duration of production activities. Predictions are contextualized within historical workflows, linking delays to prior bottlenecks or resource shortages. XAI techniques - including Shapley-based approaches \cite{lundberg2017unified}, Individual Conditional Expectation \cite{goldstein2015peeking}, and Partial Dependence Plots \cite{friedman2001greedy} - produce multiple layers of interpretability \cite{mehdiyev2025integrating}. However, grounding alone is insufficient. To foster trust, the framework embeds UQ at its core, transforming probabilistic ML outputs into risk-aware decision boundaries. UQ methods (such as Monte Carlo Dropout \cite{gal_dropout} or Split Conformal Prediction \cite{vovk2020computationally}) quantify the reliability of these predictions. Planners no longer face a binary choice between accepting or rejecting a prediction; instead, they assess predictions or credible intervals (e.g., “Cutting time: 8.2h ± 1.1h”) alongside PM insights to make informed trade-offs \cite{mehdiyev2024communicating, mehdiyev2024quantifying}. 

Since transparency without interactivity risks perpetuating the status quo, our framework introduces Module 2: a multi-agent LLM architecture that transforms static explanations into dynamic conversations. By decomposing user queries into specialized tasks — explanation generation, scenario testing, validation — the system mirrors human problem-solving: probing assumptions, testing hypotheses, and refining conclusions. This agentic design ensures explanations adapt not just to the data, but to the planner’s evolving priorities, whether optimizing for speed, cost, or resource fairness. The LLM module is refined through prompt engineering \cite{giray2023prompt}, ensuring domain-specific accuracy and mitigating bias to provide context-aware recommendations. In addition to its prompt-driven capabilities, the LLM layer employs a RAG \cite{lewis2020retrieval} mechanism to store and retrieve previously generated or curated XAI explanations: Through vector-based indexing, the LLM ranks potentially relevant explanations according to semantic similarity, quantifies their relevance scores, and selectively incorporates them into its responses. This approach not only promotes consistency and coherence in the explanations but also ensures that domain experts receive the most pertinent information for their context.

To translate these enriched conversations into actionable insights, the framework adds Module 3. This layer presents an adaptive, context‑aware user interface that integrates dashboards, conversational widgets, and visual analytics. It renders predictions, uncertainty intervals, and process‑mining traces in an intuitive format, and guides planners through what‑if analyses. By surfacing the right information at the right granularity, the interface closes the loop between analytical depth and operational usability.

Ultimately, the framework positions PM as the backbone of a collaborative loop, where human expertise and AI predictions iteratively refine one another. Planners interrogate models through natural language, which in turn respond with process-contextualized insights, and shared understanding emerges from this dialogue. In doing so, the work moves beyond technical explainability to operational trust — a prerequisite for transforming predictive analytics from a reactive tool into a proactive partner in high-stakes manufacturing.

\subsection{Experiences}
Early experiments suggest that combining robust data pipelines with iterative user interfaces significantly enhances trust among domain experts \cite{mehdiyev2024communicating, mehdiyev2024counterfactual,mehdiyev2024quantifying,mehdiyev2025augmenting,mehdiyev2025integrating}. Providing a dialogue-centric LLM interface encourages users from various domains to probe model outputs, interpret results more effectively, and adapt strategies on-the-fly. Moreover, from a data protection and compliance perspective, the LLM module supports the use of tools as well as systematic filtering and revision of outputs to ensure adherence to regulatory requirements.

While the agentic LLM architecture addresses many domain-specific queries with high accuracy, ensuring model reliability under real-time production constraints remains a challenge. The complexity of industrial processes can lead to irregularities in data quality, requiring continual data curation and pipeline updates. Additionally, harmonizing automated explanation outputs with diverse user expectations demands a careful balance between technical detail and domain-level interpretability. Finally, overcoming skepticism about “black-box” AI systems calls for ongoing transparency measures, where the interplay of XAI and LLM-driven explanations must be demonstrably fair, unbiased, and consistent with operational requirements.
\section{Use Case 2 - Conversational BPMN Modeling}
\label{section_use_case_2}

\subsection{Motivation}
Business process models are a highly valuable source of information in business contexts \cite{Leopold2013}. They are necessary for the adequate design of information systems, requirements engineering, and for the means of process communication. 
%process discovery. 
However, creating these models requires significant manual effort and knowledge about modeling notations such as BPMN. At the same time, process modeling involves various stakeholders with different levels of expertise \cite{Dumas2018_BPM}. For example, a domain expert has knowledge about the underlying business workflows; however, they might not be able to create or understand a process model on their own \cite{Meitz2013}.
Therefore, it would be beneficial to enable process modeling through natural language while eliminating the requirement of understanding the corresponding modeling notation. Furthermore, being able to ask questions about a certain process model could also improve understanding for all stakeholders.

For many years, researchers have explored automated text-to-model approaches, ranging from rule-based methods to more advanced systems using Natural Language Processing (NLP) techniques. LLMs have emerged as a promising technology for this task, given their strong reasoning capabilities over natural language and ability to solve tasks not explicitly trained on. Thus, LLMs can also support conversational process modeling, allowing for interactive, collaborative process modeling with users in a chat-based environment \cite{conv_modeling}.

A key challenge in this context is ensuring that LLMs can accurately interpret and generate BPMNs. Also, one needs to decide how the process models can be represented in a machine-readable format. We noticed that most open-source LLMs struggle to generate valid BPMN models, even though it's a frequently used modeling notation \cite{lauer_emisa_2025}. This calls for specializing LLMs for this task, e.g., by fine-tuning (preferably lightweight) LLMs.
%Therefore, fine-tuned LLMs, potentially smaller in size, are more suitable. 
In particular, Supervised Fine-Tuning on the text-to-BPMN task can enhance their ability to generate valid and accurate BPMN models.

\subsection{Approach}
%\paragraph{LLM}
We implemented a prototypical chat environment that enables users to interact with an LLM in order to generate BPMN models of business processes. Users can prompt the LLM to generate a BPMN, modify an existing BPMN, or ask clarifying questions. This interactive system requires the LLM to understand different user intents, such as initial modeling, refinement, and updates. 
To guide the LLM, a system prompt first defines the general setting and task at hand. It also defines the LLM's target output — in our case, a JSON object that can easily be parsed. It contains the textual response as well as the actual BPMN model. Additionally, one-shot prompting is used to provide an example output, which tends to improve response quality. The LLM then processes user requests, generates a new BPMN, updates an existing one, or answers questions about a BPMN model as needed. 

\begin{figure}[t]
  \centering
  \includegraphics[width=\linewidth]{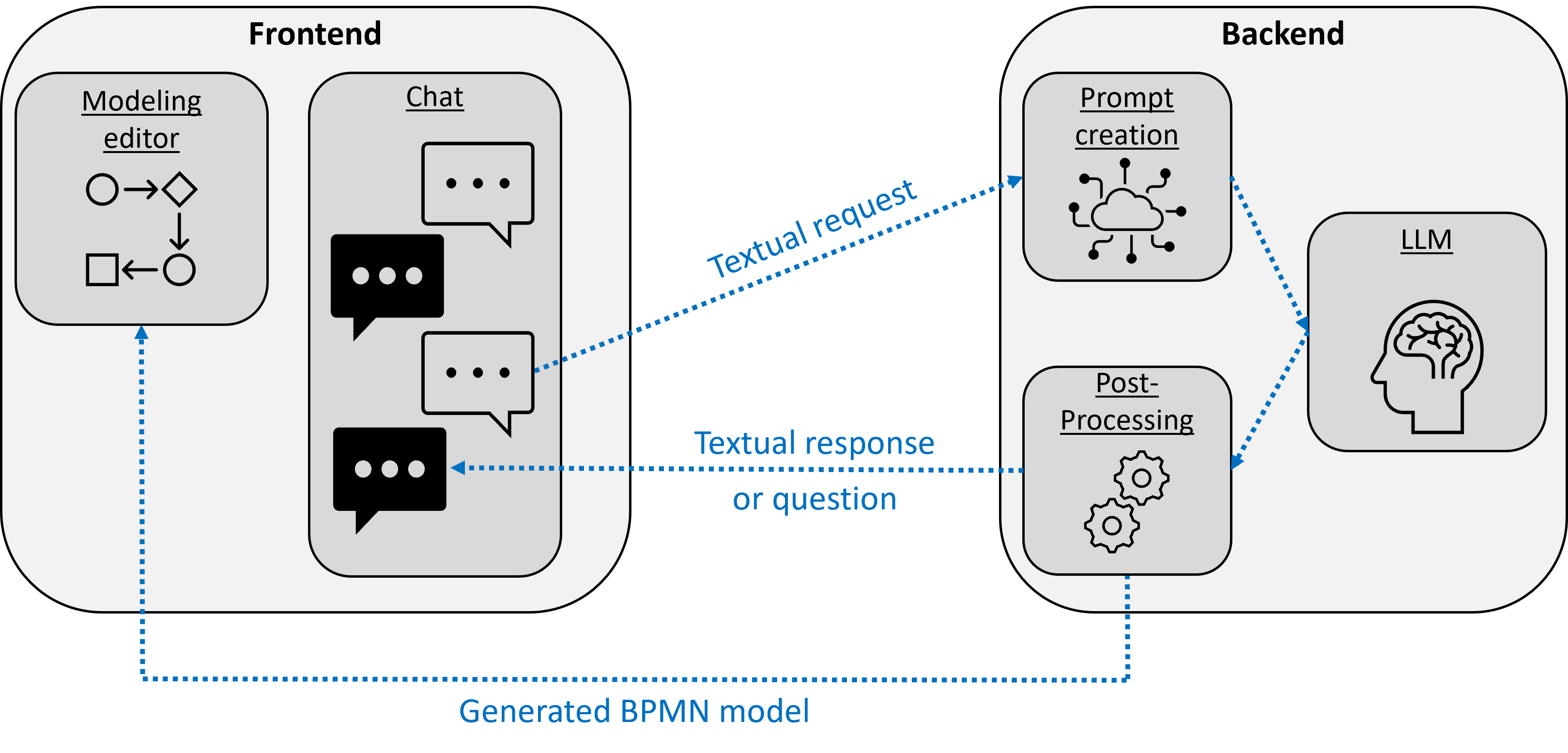}
  \caption{Architecture for Conversational Process Modeling}
  \Description{Architecture for Conversational Process Modeling}
  \label{figure_use_case_2}
\end{figure}

In our scenario, the exchange format for BPMNs between the application and LLM was defined as the official BPMN-XML standard\footnote{\url{https://www.bpmn.org/}}. To optimize efficiency, the layout of BPMN elements (i.e., precise coordinates) is handled algorithmically. This approach speeds up inference by eliminating the need for the LLM to generate layout-related tokens. At the same time, it prevents placement errors during the layouting procedure resulting from incorrect LLM outputs. 
Finally, a refinement loop was added to ensure BPMN validity, allowing the system to correct any issues before finalizing the model and presenting it to the user. The overall architecture and user collaboration is shown in \autoref{figure_use_case_2}.

%\paragraph{Evaluation}
%In preliminary experiments, we tested different LLMs with various sizes, namely Llama-3.2-1B-Instruct, Llama-3.3-70, Qwen2.5-7B-Instruct and DeepSeek-R1-Distill-Qwen-14B, to assess their performance in generating BPMN models. To this end, we gathered data from publicly available BPMN datasets, resulting in a collection of 105 samples, each consisting of a textual description as input and the corresponding BPMN XML as the target output.
In preliminary experiments, we tested various open-source LLMs with sizes between 1B and 70B on publicly available text-model pairs \cite{lauer_emisa_2025}.
To ensure the validity of the generated BPMN-XML models, we used an XML validator\footnote{\url{https://github.com/bpmn-io/bpmn-moddle/tree/main/resources/bpmn/xsd}}.
We assess the quality of generated BPMN models based on different evaluation metrics, covering different modeling aspects, such as syntactic correctness or semantic expressiveness, compared to the ground truth models.
%We assess the quality of generated BPMN models based on the SIQ-framework \cite{Reijers_Bsuiness_2010} along three metrics: syntactic (compliance with modeling rules), pragmatic quality (understandability based on 15 metrics, e.g., size, density, and concurrency \cite{Mendling_Metrics_2008}), and semantic quality (similarity to the ground truth model using 7 metrics, including natural language, graph-based, and behavioral comparisons).

\subsection{Experiences}
At first, the quality of the LLM-generated BPMNs seems convincing. The LLMs are able to generate BPMN models that depict the description, with bigger LLMs generating BPMNs with more details compared to smaller ones. However, we observe various limitations throughout our experiments (see \cite{lauer_emisa_2025} for more details).

Our findings reveal that LLMs struggle with BPMN-XML as exchange format. The XML standard requires many characters to represent a process model, such as specific XML tags that define the XML schema. Therefore, the majority of generated tokens are spent on generating content that does not address the process model itself, but rather necessary structuring elements. 
Generating BPMN-XML is also error-prone and results in many invalid models, even though a refinement loop was added. Particularly, smaller LLMs struggled to produce valid and syntactically correct BPMNs. In contrast, larger LLMs performed noticeably better, as expected.

Further, the BPMNs have semantic deficits. Often, the LLMs struggle to identify all the relevant behaviors of a business process and thus generate BPMNs that do not contain all paths. This means that either a manual refinement or additional instructions on how to update specific details of the BPMN are necessary. We also observed that when prompted to alter an existing BPMN, LLMs often failed to incorporate all necessary changes, especially in terms of updating, removing, or adding the right sequence flows. 

When asked to generate complex BPMNs, LLMs tend to produce faulty code or hallucinate non-existent XML tags. Therefore, it is advisable to introduce a more compact process model representation. In literature, different approaches are used, e.g., utilizing bullet point lists \cite{10.1007/978-3-031-50974-2_34} or partially ordered workflow language, which can be converted to process models \cite{10.1007/978-3-031-61007-3_18}. Further, Chain-of-Thought should be used to increase the accuracy of the generated models and to make the modeling intention of the LLM transparent to the user, fostering trust in the LLM and understandability of the BPMN.

Additionally, inference times were quite long, making larger LLMs less suitable for real-time, chat-based interactions. This highlights the need for lightweight LLMs for this use case to overcome scalability issues.
Fine-tuning via techniques such as LoRA \cite{lora} can be an efficient way to enhance LLM performance in conversational process modeling. However, creating high-quality fine-tuning datasets can be challenging and time-consuming.
So-called Preference Tuning can also be used to let the LLM obtain a deeper understanding of process modeling. For example, one could provide a textual process description and different corresponding process models - in this scenario BPMNs - with different levels of quality and let the LLM decide which process model fits best. 
Coming from the other direction, one could provide a process model as well as different textual descriptions that describe this process more or less well, and let the LLM decide which description is the best fitting one. In real-world settings, it's possible to model a certain process in different correct ways. It is precisely this aspect that could be trained with such an approach, where the LLM learns nuances in business process modeling. However, this also requires high-quality preference datasets as well as additional fine-tuning stages (e.g., leveraging Direct Preference Optimization \cite{dpo}).

%We found that a significant portion of the generated BPMNs had syntactic errors, with smaller LLMs particularly struggling to produce valid BPMNs. In contrast, larger LLMs performed noticeably better, as expected. Also, none of the tested LLMs were able to generate semantically sound models. Often, they struggle to identify all the relevant behavior of a business process and thus generating BPMNs that do not contain all paths. This means that either a manual refinement is necessary or that its required to prompt the LLM with specific details on how to update the BPMN. 
%However, we observed that when prompted to alter an existing BPMNs, LLMs often failed to incorporate all necessary changes, especially in terms of updating, removing, or adding sequence flows. 

%Additionally, inference times were quite long, making larger LLMs less suitable for real-time, chat-based interactions. This highlights the need for lightweight LLMs for this use case.
%Fine-tuning via techniques such as LoRA \cite{lora} can be an efficient way to enhance LLM performance in conversational process modeling. However, creating high-quality supervised fine-tuning (SFT) datasets can be challenging and time-consuming. Preference optimization also has the potential to let the LLM obtain an understanding of different levels of quality regarding process modeling. However, this also requires high-quality preference datasets.

\section{Use Case 3 - Monitoring Drug Safety with KGs and LLM-Agents}
\label{section_use_case_3}
\subsection{Motivation}
The safety of medicines has to be monitored continuously throughout their use. Pharmacovigilance (PV) describes the activities carried out to ensure the safety of medicines in the market, i.e., the detection, assessment, understanding, and prevention of adverse effects with medicines. 
The PV lifecycle consists of the phases ''collect, manage and analyze'', ''review and assess'', ''decide and act'', ''communicate'', and ''monitor'' \cite{EMA2021}. Information on medicines in the market has to be collected and analyzed, reviewed, and assessed by regulatory experts to decide whether and which actions are required. These actions need to be communicated and their consequences monitored.

These tasks require screening large amounts of textual reports and their assessment with domain knowledge. For instance, they involve querying and scanning relevant documents in public databases such as PubMed (scientific reports) and MAUDE (reports from medical experts and patients). Further, internal company case reports or social media posts can also be of interest. Domain knowledge in the form of publicly\footnote{e.g., \url{https://bioportal.bioontology.org/}} or internally available KGs is another relevant source of information, such as drug-drug interactions or terminologies. This requires querying graph databases. In order to analyze and decide about following actions in PV, experts need to interpret information with vast domain knowledge, e.g., to assess whether a reported issue is known or unknown, or even an adverse event.

LLMs' abilities can assist regulatory experts with PV. LLMs with tool access can retrieve, collect, and analyze reports from different data sources. Graph-structured databases can also be queried by LLMs \cite{sequeda2024benchmark}, and the information linked with reports. This supports regulatory experts in the initial phases of the PV lifecycle. Additionally, LLM agents can support other lifecycle phases, like ''communicate'' and ''monitor'' by writing reports and using agents to perform the data collection phase continuously. Thereby, they support experts by relieving them from repetitive and time-consuming tasks such as information retrieval and initial analysis. 
However, data is distributed across various databases and stored in different formats such as text and graphs. Further, strict regulations apply when integrating AI systems into highly regulated environments. Therefore, a custom approach has been designed.

\begin{figure}[t]
  \centering
  \includegraphics[width=\linewidth]{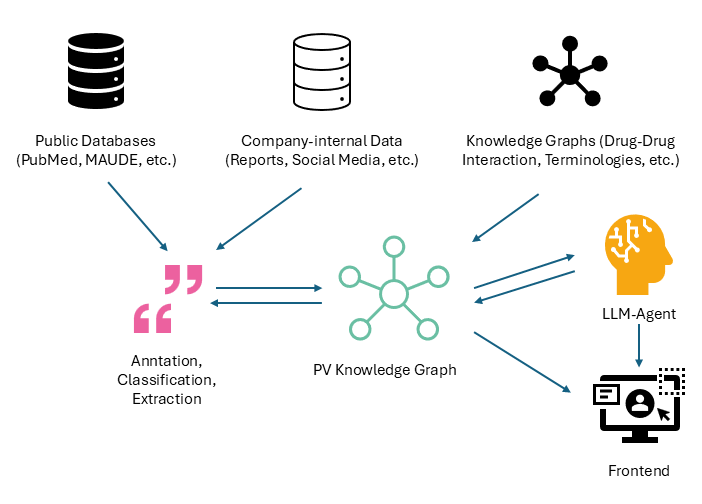}
  \caption{Architecture of PV KG and LLM-Agents}
  \label{figure_use_case_3}
  \Description{Conceptual Framework - Use Case 3}
\end{figure}

\subsection{Approach}
The main parts of the approach are outlined in \autoref{figure_use_case_3}, consisting of database extractors (top row), NLP preprocessing, the PV KG, LLM agents, and a frontend to interact with users. Central is a custom PV KG containing entities and relations relevant to PV. Based on this, LLM agents, which are able to utilize the information in the PV KG and other KGs, are used to answer user queries and PV questions.

\paragraph{Database Extractor}
Different database extractors retrieve relevant reports and other information from diverse data sources, e.g., scientific publications from PubMed, reports from MAUDE, or social media posts. Using language models, these texts are annotated and classified, and relevant entities are extracted, which allows information to be stored in a KG. If data is readily available as a graph, e.g., publicly available KGs, we link or integrate this data directly into the PV KG (e.g., terminologies like SNOMEDCT or RXNORM).

\paragraph{PV KG}
A dedicated KG for PV is created, consisting of relevant information in a structured format, such as drug-drug interactions, adverse effects, populations, risks, and subsets of terminologies. Further, it should link to the source document, e.g., the PubMed report, social media post, or other KG, if detailed information is required. As such, it serves as a unified knowledge base from which to retrieve information.

\paragraph{LLM Agents}
LLMs with tool access are used to retrieve information from the PV KG for analytical tasks. For this, LLM-based question answering over KG is employed \cite{sequeda2024benchmark}, where the LLM queries the KG given a question from the user and answers that question based on the results from the KG query. By using structured knowledge in KG, resembling domain knowledge, the answers from the LLM should adhere to factual knowledge instead of only learned knowledge, reducing the risk of hallucinations \cite{wu2025graph}. 
LLM agents will be used for more complex tasks that require multiple steps, e.g., collecting data from databases that cannot be queried or more complex queries that require analyzing multiple reports and documents in depth. They are trained from recorded process executions by users to perform the same task similarly. Individual steps can be performed by LLMs with tool access. Thereby, they support humans in information retrieval and analysis.

%Muss hier stehen bleiben da die Fig sonst auf die letzte Seite rutscht
%\begin{figure*}[!t]
%  \centering
%  \includegraphics[width=0.92\linewidth]{images/Use Case 4 - Framework.png}
%  \caption{Architecture of Multi-Agent Textile Design Assistance System}
%  \Description{Conceptual Framework of Use Case 4}
%  \label{figure_use_case_4}
%\end{figure*}

\paragraph{Frontend}
The frontend is used for displaying results and interacting with the databases and agents. It allows the user to manually query the databases, in case they want to analyze the data by hand. It also allows to trigger agentic workflows.

\subsection{Experiences}
Preliminary results indicate the overall approach's feasibility. For instance, it is feasible to retrieve, annotate, and extract entities of interest from PubMed and MAUDE and store them in the PV KG using language models. Small-sized LLMs are sufficient for most tasks, while larger LLMs are beneficial for complex reports.
Additionally, LLMs are capable of formulating queries to the PV KG given the respective KG scheme, as well as interpreting the output. This should allow multi-step pipelines where the LLM decides to query the KG and/or other documents to extract relevant information.
Next steps include validating the quality of the generated answers as well as the performance and accuracy of the results of the agentic workflows. Multi-step LLM agent workflows will be conceptualized and trained using the recorded process executions to validate the feasibility of instructing agents with click stream data. We expect the agents to take over a substantial amount of repetitive work with some applications on the analysis part, too. Further, a framework for evaluating the compliance of the approach has to be developed.
\section{Use Case 4 - Multi-Agent Assistance System for the Textile Design Process}
\label{section_use_case_4}

\begin{figure*}[!t]
  \centering
  \includegraphics[width=0.85\linewidth]{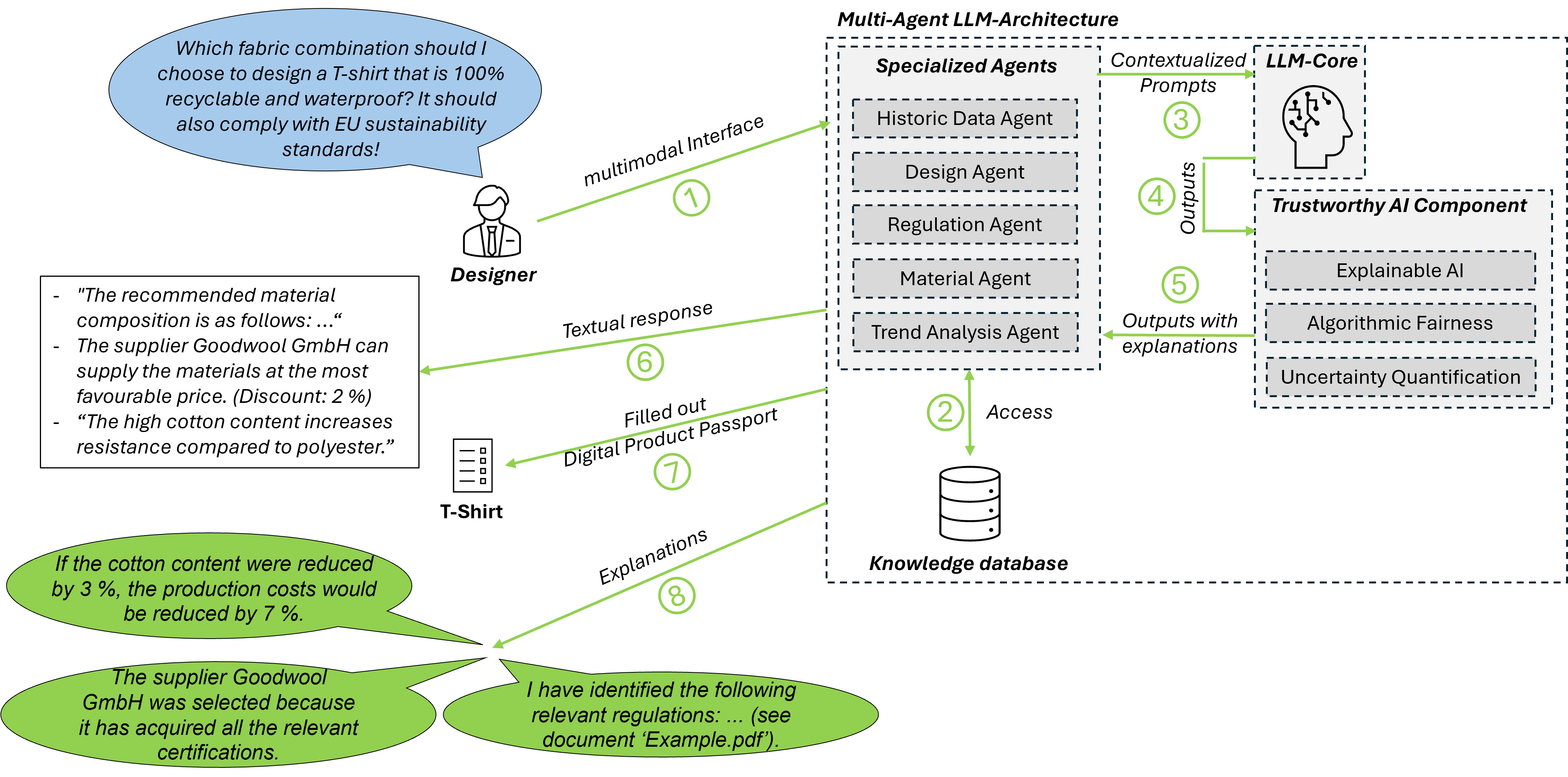}
  \caption{Architecture of Multi-Agent Textile Design Assistance System}
  \Description{Conceptual Framework of Use Case 4}
  \label{figure_use_case_4}
\end{figure*}

\subsection{Motivation}
The textile industry is one of the most significant global economic sectors and faces various challenges such as environmental concerns, resource inefficiency, waste generation, and the growing pressure to meet regulatory and sustainability standards \cite{EU_textiles_QA, textile_slr}.
Increasing consumer demand for environmentally responsible products and more stringent regulations have created a shift in the textile industry towards sustainability and circular economy principles. However, this transformation is complex, requiring innovative solutions from the sourcing of raw materials to managing the end-of-life of textiles \cite{textile_slr,textile_slr2}.

In response to these challenges, the EU has launched a strategy for sustainable and circular textiles \cite{EUROPEANCOMMISSION2022}. The strategy aims to promote sustainable practices, reduce environmental impact, and increase the efficiency of material usage throughout the textile supply chain. For instance, the Digital Product Passport (DPP) aims to enhance transparency throughout a fashion product's lifecycle.
However, operationalizing this strategy is challenging. It requires industry-wide collaboration, significant technological advances, and the development of reliable decision-making frameworks based on data about materials, production processes, and supply chains. As a result, there is a pressing need for more intelligent, data-driven approaches to help manufacturers meet these challenges.

\subsection{Approach}
The goal is to implement a trustworthy AI system that processes heterogeneous data such as DIN Standards, EU guidelines, or material characteristics, to support textile designers during the design process. The system shall facilitate adherence to regulatory requirements and actively allow for sustainable textile designs. It will assist designers by providing actionable recommendations for sustainable practices across multiple design criteria.

For instance, the longevity of textiles can be enhanced by using durable fabrics. However, sometimes, a tradeoff has to be made between durability and recyclability, since not all durable fabrics can be recycled easily or infinitely. Similarly, tradeoffs between social and economic factors have to be made when choosing suppliers. The proposed system should support the designer in making informed decisions that align with sustainability goals. By analyzing the specific context of each design decision, the system will also generate explanations to ensure transparency and reliability.
Further, DPPs should be created automatically by the system, thus reducing the designers' need for manual work.
%The proposed AI system will automate the creation of DPPs and thus relieve the designer of a lot of manual work. Within the BPM lifecycle, the approach is relevant for the \textit{Process execution} phase.

The core of the solution will be a multi-agent LLM system to facilitate the interaction between designers and sustainability criteria. The first key component consists of specialized LLM-agents, as shown in \autoref{figure_use_case_4}, that each perform a specialized task. For example, the regulation agent searches the knowledge base for relevant data on regulations and laws in order to integrate them into the design process. The material agent specializes in identifying materials and suppliers following the requirements requested by the designer. The underlying knowledge base includes various internal and external data sources. The user can upload company-owned data, such as historic orders or internal sustainability criteria. This shall allow the system to access relevant data at all times and ensure that the final results are tailored to the end-user.
All agents are controlled by a central orchestrator, which controls the information flow between the agents to ensure an adequate execution.

Given the results of the specialized agents, the LLM-core is used to generate corresponding outputs. To build confidence in the system’s recommendations, a key component will be trustworthy AI. The idea is to leverage XAI techniques such as counterfactual explanations, or UQ approaches like Bayesian approaches or Conformal Prediction. These methods are built on top of the LLMs and are used to analyze their outputs, which would otherwise be opaque from a human point of view. The final outputs and explanations will then be forwarded to the designer.
This approach will allow for an easy integration of textual data and contextual inputs from various sources. Designers will be able to interact with the system to quickly generate ideas, validate choices, and access data-driven insights. 
%As mentioned before, the system will also assist in filling out the DPP.

\subsection{Experiences}
We started implementing this use case by developing a recommender system for fabrics based on user preferences. Unfortunately, there is very little data on fabric combinations and their longevity publicly available, which is why expert interviews have been conducted. Thereby, we noticed that several trade-offs have to be made, which requires extensive explanation and communication with the designer about the consequences of substituting fabrics, other implications, e.g., social aspects, and alternatives.
%in order to inform them about the possibilities of using other fabrics, their implications, and alternatives.
Soon, the impact of suppliers and their certificates became evident, which adds another dimension to the decision problem. Using LLMs as well as Trustworthy AI techniques should allow the unaware end-user to understand this multi-criteria decision problem and balance the effects on each dimension. For example, the system could identify that selecting a specific supplier for a specific material could reduce the production cost by a considerable amount while still adhering to the same sustainability criteria as before. 
There is an ongoing challenge on how to adequately integrate corresponding knowledge into the LLM system and how to make those trade-offs explainable to the end-user by means of XAI and related methods.

\section{Discussion}
\label{section_discussion}

\begin{table*}[!t]
\centering
\caption{Comparative Analysis of LLM-Driven BPM Use Cases}
\label{tab:comparison}
\begin{tabular}{lp{3.2cm}p{3.2cm}p{3.2cm}p{3.2cm}}
\toprule
\textbf{Use Case} & \textbf{LLM-based Interaction with Predictions} & \textbf{Conversational BPMN Modeling} & \textbf{Pharmacovigilance with AI Agents} & \textbf{Sustainable Textile Design Process} \\ 
\midrule
\textbf{Input Modality} & 
Event logs & 
Text & 
Text, KG & 
Text, Tabular, Images \\ \midrule

\textbf{BPM Lifecycle Phases} & 
Implementation, Execution, Monitoring & 
Discovery, Redesign & 
Implementation, Monitoring & 
Execution \\ \midrule

\textbf{Industry Impact} & 
Transparent, process-driven production planning & 
Democratized process design (cross-industry) & 
Supporting regulatory experts in medicine safety & 
Sustainable textile design \\ \midrule

\textbf{LLM Role} & 
Multi-agent dialogues with Trustworthy AI-based PPM  & 
Conversational BPMN translation & 
KG-QA agents for pharmacovigilance & 
Multi-agent orchestration for trade-off resolution \\ \midrule

\textbf{Trust Mechanism} & 
UQ, XAI, RAG & 
SFT, Chain-of-Thought & 
KG grounding & 
XAI, Algorithmic Fairness, RAG \\ \midrule

\textbf{Ethical Risks} & 
Accountability in high-stakes decisions & 
Over-reliance on automated modeling & 
Hallucinations in medical compliance & 
Bias in sustainability trade-offs \\ \midrule

\textbf{Human-AI Interaction} & 
Multi-agent dialogues with process-contextualized feedback & 
Natural language dialogue & 
Agentic workflows with human oversight & 
Committee-based multi-agent negotiation \\ \midrule

\textbf{Key Challenges} & 
Scalability& 
Improving BPMN quality & 
LLM-KG integration & 
Quantifying multidimensional trade-offs \\ \midrule

\textbf{Data Challenge} & 
Event log complexity & 
Efficient BPMN notation modeling & 
Multisource integration (text + KG) & 
Multimodal integration (materials, regulations) \\ \midrule

\textbf{Future Directions} & 
Bilateral integration of UQ and XAI; Object-centric PM & 
Lightweight LLMs for real-time BPMN modeling & 
Hybrid KG-LLM architectures & 
Multi-objective optimization frameworks \\
\bottomrule
\end{tabular}
\end{table*}

\subsection{Overview}
\autoref{tab:comparison} summarizes and compares the four use cases in terms of input modality, BPM lifecycle phases, industry impact, LLM Role, XAI perspectives, challenges, and future directions. As demonstrated throughout the paper, LLMs can assist various BPM use cases in different industries - from production to consulting and life-science to sustainable textiles.
All BPM lifecycle phases can be supported with LLM techniques, impacting different business goals such as efficiency, compliance, and sustainability. To enforce trust in AI-driven BPM solutions, various trust mechanisms such as Shapley values, Chain-of-Thought, or KG grounding can be used. In all use cases, human interaction is enforced, and no decisions are made without human supervision.

\subsection{Challenges}
Each use case has its own challenges. In the first use case, generating real-time UQ and XAI outcomes in industrial PPM is a key challenge. Explanations have to balance technical details and domain-level interpretability to be trusted by the users.

In collaborative BPMN modeling, fine-tuning small LLMs to generate high-quality BPMN models in near real-time is of particular interest. The lack of high-quality datasets for this task makes training LLMs particularly challenging. Therefore, alternative training strategies have to be developed to enhance the modeling competencies of open-source LLMs.

To support regulatory experts in PV, challenges remain in combining LLMs with KGs. Especially when multiple KGs are available or multiple steps are required to reach conclusions. Instructing agents to perform business processes following human demonstrations will be another challenge to face.

For sustainable recommendations in textile designs, trade-offs in multiple dimensions and domains, which are difficult to quantify, have to be made. Whether LLMs can balance these trade-offs remains to be seen. 

Spanning across all use cases, we see that more research is needed for process-aware LLMs. While LLMs can generate text that is difficult to distinguish from expert writing, LLM-generated BPMN models are often not valid XML files and contain syntactical errors. Further, agentic workflows need instructions, showing that such systems are not process-aware by design. However, we see the potential to transfer knowledge between the use cases to enhance the processual thinking of LLMs.

Challenges also remain when it comes to ethical risks. Over-reliance on predictions or generated BPMN models, which can potentially be faulty, as well as biases in LLMs, poses the risk of leading to bad consequences. For regulatory use cases, hallucinations have to be controlled, and the LLMs have to be enforced to only use factual knowledge available in KGs.

\begin{comment}

\begin{table*}[htbp]
    \centering
    \begin{tabular}{lllll}
        \toprule
        Use Case & Domain & Input Modality & Lifecycle Phases & Techniques \\ 
        %\midrule
        XAI PPM & Manufacturing & ? & ? & ? \\
        Process modeling & Consulting & Text                & Discovery, Redesign & LLM \\
        Pharmacovigilance & Life-science & Text \& KG       & Implementation, Monitoring & LLM, Agents/Tooling, RAG \\
        Clothing design & Textile & Text, tabular, image    & ? & LLM, Agents, RAG \\
         \bottomrule
    \end{tabular}
    \caption{Overview of presented use cases}
    \label{tab_use_cases}
\end{table*}

\end{comment}
\section{Conclusion}
\label{sec:conclusion}
This paper introduced four LLM-based use cases, demonstrating that all phases of the BPM lifecycle can be supported. These cases span different industries, such as manufacturing, consulting, life-science, and sustainable textiles. They support companies by increasing efficiency, enhancing business processes, ensuring compliance, and adhering to diverse regulations. Various ways on how to apply LLMs in productive settings have been shown - from explaining process predictions, generating BPMN models, querying knowledge graphs to balancing trade-offs.
In all cases, it emerged that human-machine interactions are essential for successful process executions. In addition, AI methods - and LLM applications in particular - should be supplemented by XAI techniques in order to make their predictions auditable for decision-making in real-world scenarios.
Key challenges across the use cases remain in improving LLMs' capabilities to think in processes rather than support single tasks. Further, trustworthiness should be enforced by design, and ethical constraints should be considered from early on.


\begin{acks}
This research was funded in part by the German Federal Ministry of Education and Research under grant number 01IS24048C (EINHORN), 01IS24053C (KICoPro), and 01IS24046D (PVRadar).
\end{acks}

\bibliographystyle{ACM-Reference-Format}
\bibliography{sample-base}

%\appendix
%\section{Appendix A}

\end{document}